\documentclass[epj]{webofc}
\usepackage[varg]{txfonts}   % Web of Conferences font
\newcommand{\DZero}{\rm D\O}
\def\MeV{\mbox{ MeV}} 
\def\GeV{\mbox{ GeV}} 
\def\beq{\begin{equation}}
\def\enq{\end{equation}}
\def\beqa{\begin{eqnarray}}
\def\enqa{\end{eqnarray}}

\newcommand{\rag}{\rangle}
\newcommand{\lag}{\langle}
\def\lb{\label}
\def\qq{\lag\bar{q}q\rag}

\wocname{EPJ Web of Conferences}
\woctitle{CONF12}
\begin{document}
\selectlanguage{english}
\title{$X,~Y$ and $Z$ States}
%
% subtitle (optional, strongly discouraged)

\author{M.~Nielsen\inst{1}\fnsep\thanks{\email{mnielsen@if.usp.br}}, 
%\and
R.~M. Albuquerque\inst{2},
J.~M. Dias\inst{1},
K.~P. Khemchandani\inst{2}, 
A.~Mart\'inez~Torres\inst{1},
F.~S. Navarra\inst{1}\and
C.~M. Zanetti\inst{2}
}

\institute{Instituto   de  F\'{\i}sica, 
  Universidade de S\~{a}o Paulo, C.P.  66318, 05314-970 S\~{a}o Paulo, 
  SP,    Brazil
\and
Faculdade de Tecnologia, Universidade do Estado do Rio de 
Janeiro, Rod. Presidente Dutra Km 298, P\'olo Industrial, 27537-000 , 
Resende, RJ, Brasil 
}

\abstract{%
  Many new states in the charmonium mass region were recently
discovered  by  BaBar, Belle, CLEO-c, CDF, \DZero, BESIII, LHCb and CMS 
Collaborations. We use the QCD Sum Rule approach to study the possible 
structure of some of  these states. 
}
\maketitle
\section{Introduction}
\label{intro}
During the last decade several experimental facilities, such as BaBar 
at SLAC and Belle at KEK, CLEO-III and CLEO-c at CESR, 
CDF and \DZero\ at Fermilab,  BESIII at IHEP and LHCb and CMS at CERN, 
have increased the available data on new charmonium-like states, called 
$X,~Y$ and $Z$ states. Tables with a list  of these  states can be 
found in Refs.~\cite{Brambilla:2014jmp,Hosaka:2016pey}. In 2014 there were
twenty three of these $X,~Y,~Z$ states, many not confirmed. They do not 
seem to have a simple $c\bar{c}$ structure.  Although the masses of these
states are above the corresponding thresholds of  decays into a pair of 
open charm mesons, they decay into $J/\psi$ or $\psi^\prime$ plus pions, 
which is unusual for $c\bar{c}$ states. Besides, their masses and decay 
modes are not in agreement with the predictions of potential models, which, 
in general, describe very well $c\bar{c}$ states. For these reasons they are 
considered as being good candidates of exotic hadrons. We call exotic 
states hadrons having other structure than the ordinary mesons and baryons, 
containing constituent quark-antiquark and three quarks respectively. 
The idea of unconventional quark structures is quite old and the light 
scalar mesons were the first candidates for tetraquark exotic states. 
These states are allowed by the strong interactions, both at the fundamental
level and at  the effective level, and their absence in the experimentally 
measured spectrum has  always been  a mystery.

Among these new charmonium states, the charged ones are definitely exotics, since 
they can not be simple $c\bar{c}$ states. The $Z^+(4430)$, found by the Belle 
Collaboration in 2007, was the first observed one 
\cite{Choi:2007wga,Mizuk:2009da,bellez2}. 
The BaBar Collaboration searched for the $Z^-(4430)$ signature in
four decay modes and concluded that there is no significant evidence for a 
signal peak  in any of these processes \cite{babarz}. However, a few years ago 
Belle  and LHCb collaborations have confirmed the $Z^+(4430)$ observation 
and have determined the  preferred assignment of the quantum numbers to be 
$J^{P} = 1^{+}$ \cite{bellez2,Aaij:2014jqa}. The LHCb Collaboration also did the 
first attempt to demonstrate the resonant behavior of the $Z^+(4430)$ 
state ~\cite{Aaij:2014jqa}. They have performed a fit in which the Breit-Wigner 
amplitude was replaced by a combination of independent complex amplitudes at six 
equally spaced points in $m_{\psi(2S)\pi}$ range covering the $Z^+(4430)$ peak 
region. The resulting Argand diagram is consistent with a rapid phase 
transition at the peak of the amplitude, just as expected for a resonance.   
Therefore, the confirmation of the observation of the $Z^+(4430)$ by the LHCb 
Collaboration  with the demonstration of its  resonant behavior
can be considered as the first experimental proof of  the 
existence of the exotic states.  

The other
very interesting state, which is the most well studied among the new 
charmonium states, is the $X(3872)$. It was first observed  in 2003 by the  Belle 
Collaboration \cite{Choi:2003ue,Adachi:2008te},  and has been confirmed by  
five collaborations: BaBar~\cite{Aubert:2008gu}, CDF~
\cite{Acosta:2003zx,Abulencia:2006ma,Aaltonen:2009vj}, 
\DZero~\cite{Abazov:2004kp}, LHCb~\cite{Aaij:2011sn,Aaij:2013zoa}
and  CMS~\cite{Chatrchyan:2013cld}. 
The LHCb collaboration has determined the 
$X(3872)$ quantum numbers to be $J^{PC} = 1^{++}$, with more than 8$\sigma$ 
significance \cite{Aaij:2013zoa}. 
Calculations using constituent quark models 
give masses for  possible charmonium states, with $J^{PC}=1^{++}$ quantum 
numbers,  which are much bigger than the observed $X(3872)$ mass: 
$2~^3P_1(3990)$ and $3~^3P_1(4290)$ \cite{bg}. These results, together with 
the coincidence between the $X$ mass and the $D^{*0}D^0$ threshold: 
$M(D^{*0}D^0)=(3871.81\pm0.36)\MeV$ \cite{pdg}, inspired the proposal that 
the $X(3872)$ could be a molecular $(D^{*0}\bar{D}^0+\bar{D}^{*0}D^0)$ bound 
state with a small binding energy \cite{swanson}.
Other interesting possible interpretation of the $X(3872)$, first proposed 
in Ref.~\cite{maiani}, is that it could be a tetraquark state
resulting from the binding of a diquark and an antidiquark. The difference 
between these two interpretations is only the way that the 4-quarks are 
organized 
inside the state. A molecular $(D^{*0}\bar{D}^0+\bar{D}^{*0}D^0)$ 
bound state with small binding energy would be bigger than a compact tetraquark
state. In any case, there is little doubt in the
community that the $X(3872)$ structure is more complex than  just a  
$c\bar{c}$ state. 

In the following we discuss some of these new states using the QCD sum rule 
(QCDSR) approach.

\section{QCD Sum Rules}

The method of the QCDSR, was introduced by Shifman, Vainshtein and 
Zakharov \cite{svz} for the study of the mesons. They demonstrated that, for 
the determination of the 
mass of the state using the method, the non-perturbative power corrections
are more important than the strong coupling, $\alpha_s$, corrections.
The non-perturbative power corrections were introduced through a series 
expansion of operators. As the dimension of the operators increase,
the power of the momentum transfer, $Q^2$, in the denominator
of the terms also increases, giving a series in $1/Q^2$ which can be 
truncated for large values of $Q^2$. The sum rule
method was latter extended to baryons by Ioffe \cite{io1} and Chung 
{\it et al.} \cite{dosch}. Since then the QCDSR technique has been applied to 
study numerous hadronic properties with various flavor content and has been 
discussed in many reviews \cite{rry,SNB,SNB2,col,review,Nielsen:2014mva}
emphasizing different aspects of the method.

%%%%%%%%%%%%%%%%%%%%%%%%%%%%%%%%%%%%%%%%
\begin{table}[h]
\begin{center}
\caption{Charmonium states observed in the last years.}
\label{tabnew}       
\begin{tabular}{|c|c|c|c|} \hline
      state    & Production mode  & Ref.  \\ \hline
  $X(3872)$ & $B\to K (\pi^+\pi^-J/\psi)$&  \cite{Choi:2003ue}  \\
  $Y(4260)$ & $e^+e^-\to\gamma_{ISR}(J/\psi\pi^+\pi^-)$ & \cite{babar1}\\
  $Z_c^+(3900)$ & $ e^+~e^-\to \pi^-(\pi^+J/\psi)$ &\cite{Ablikim:2013mio} \\
  $Z_c^+(4025)$ & $ e^+~e^-\to \pi^-(D^*\bar{D}^*)^+$ &\cite{Ablikim:2013emm}\\
  $X^\pm(5568)$ & $p~\bar{p}\to(B_s^0\pi^\pm)+\cdots$  & \cite{D0:2016mwd}\\
\hline
     \end{tabular}  
\end{center}
\end{table}
%%%%%%%%%%%%%%%%%%%%%%%%%%%%%%%%%%%%%%%%

The method is based in the evaluation of the correlation function:
\beq
\Pi(q)\equiv i\int d^4 x\, e^{iq\cdot x}
\lag{0}| T [j(x)j^\dagger(0)]|0\rag\ ,
\label{cor}
\enq
in two different ways. At the quark level in terms of quark 
and gluon fields, and at the hadronic level by
introducing hadron parameters. In Eq.~(\ref{cor}) $j(x)$ is a current 
which has the quantum numbers of the hadron we want to study. 

In what follows we present some results of the QCDSR calculations on the 
$X,~Y,~Z$ states presented in Table \ref{tabnew}. 
We assume these states to have a more complicated structure than simple
quark-antiquark states. For more details we refer the reader to our recent 
reviews on the subject \cite{review,Nielsen:2014mva}.

\section{$X,~Y$ and $Z$ states}

\subsection{$X(3872)$}
\label{3-1}

As discussed in Sec.~\ref{intro}, 
the $X(3872)$  was the first observed non-conventional charmonium, i.e., it 
has a mass significantly smaller than the one predicted by the standard 
quark model with these quantum numbers, which are $J^{PC} = 1^{++}$
\cite{Aaij:2013zoa}. Moreover, the $X$ 
decays with comparable strength into $J/\psi$ plus two and  $J/\psi$ plus 
three pions \cite{pdg}, showing a strong isospin violation which is not 
compatible with a  $c - \bar{c}$ state.  
Finally, this state has a decay width of less than  $1.2$ MeV \cite{pdg}, 
which is too small to be easily accounted for. 

If we assume the $X$ to be described by a $J^{PC}=1^{++}$ four-quark 
current either in a diquark-antidiquark configuration:
\beq
j^{(q,di)}_\mu={i\epsilon_{abc}\epsilon_{dec}\over\sqrt{2}}[(q_a^TC
\gamma_5c_b)(\bar{q}_d\gamma_\mu C\bar{c}_e^T)
+(q_a^TC\gamma_\mu c_b)
(\bar{q}_d\gamma_5C\bar{c}_e^T)]\;,
\label{cur-di}
\enq
or in a molecular $D\bar{D}^*$ configuration:
\beq
j^{(q,mol)}_{\mu}(x)  =  {1 \over \sqrt{2}}
\bigg[
\left(\bar{q}_a(x) \gamma_{5} c_a(x)
\bar{c}_b(x) \gamma_{\mu}  q_b(x)\right) - 
\left(\bar{q}_a(x) \gamma_{\mu} c_a(x)
\bar{c}_b(x) \gamma_{5}  q_b(x)\right)
\bigg],
\lb{cur-mol}
\enq
we find that it is possible to describe its mass \cite{x3872,lnw}, but
it is not possible to describe its width \cite{decayx}.
The mass obtained using the current in  Eq.~(\ref{cur-di}) 
was $M_X=(3.92\pm0.13)~\GeV $ \cite{x3872} .  
In the case of the current in  Eq.~(\ref{cur-mol}),  the result for the mass 
obtained in Ref.~\cite{lnw} was $M_X=(3.87\pm0.07)$  GeV, both in  good
agreement with the experimental mass. However, the decay width
obtained in Ref.~\cite{decayx} for the decay mode $X\to J/\psi\pi\pi$
was $\Gamma_{X\to J/\psi\pi\pi}=(50\pm15)~\MeV$, much bigger than the 
experimental upper limit. Therefore, from a QCDSR calculation it 
is not possible to explain the small width of the $X(3872)$ if it is a pure
four-quark state. In Ref.~\cite{x24} the $X(3872)$ 
was treated as a mixture of a $c\bar{c}$ state with a four-quark state: 
\beq
J_{\mu}^{q}(x)= \sin(\alpha) j^{(q,4)}_{\mu}(x) + \cos(\alpha) 
j^{(q,2)}_{\mu}(x),
\lb{cur24}
\enq
with $j^{(q,4)}_{\mu}(x)$ given in Eq.~(\ref{cur-di}) or Eq.~(\ref{cur-mol})
and
\beq
j^{(q,2)}_{\mu}(x) = {1 \over 6 \sqrt{2}} \qq [\bar{c}_a(x) \gamma_{\mu}
\gamma_5 c_a(x)].
\enq

The necessity of mixing a $c\bar{c}$ component with a molecule was already 
pointed out in some works \cite{mechao,suzuki,dong,li1}. In particular,
in Ref.~\cite{grinstein}, a simulation of the production of a bound 
$D^0\bar{D}^{*0}$ state with binding energy as small as 0.25 MeV, obtained a
cross section of about two orders of magnitude smaller than the prompt
production cross section of the $X(3872)$ observed by the CDF Collaboration.
The authors of Ref.~\cite{grinstein} concluded that $S$-wave resonant 
scattering is unlikely to allow the formation of a loosely bound 
$D^0\bar{D}^{*0}$ molecule in high energy hadron collision.   On the other 
hand,  the CDF data on $X(3872)$ production were well explained in \cite{han} 
where the authors assumed that it is a mixture with a  two-quark 
($\chi_{c1}^{\prime}$) and a four-quark  ($D\bar{D}^*$) component.

From the results presented in Ref.~\cite{x24} one can conclude that it is 
possible to reproduce the experimental mass of the $X(3872)$ for a wide 
range of  mixing angles, $\alpha$,  but, as observed in \cite{x24}, it is  
not so easy to reproduce the experimental decay width. 
In  Ref.~\cite{x24} it was shown that
with  a mixing angle  $\alpha=9^0\pm4^0$ in Eq.~(\ref{cur24}) 
it is possible to describe the experimental mass of the $X(3872)$ with
a decay width of  $\Gamma(X\to J/\psi ~(n\pi))=(9.3\pm6.9)~\MeV$, which is  
compatible with the experimental upper limit. 

To summarize, we could say that, in a QCDSR
calculation, the $X(3872)$ can be well described basically by a
$c\bar{c}$ current with a small, but fundamental, admixture of molecular
($D\bar{D}^*$) or tetraquark ($[cq][\bar{c}\bar{q}]$) components \cite{x24}. 

\subsection{$Y(4260)$ }

The $Y(4260)$ was first observed by the BaBar collaboration in the $e^+e^-$ 
annihilation through initial state radiation \cite{babar1}, and it was 
confirmed by the  CLEO and Belle collaborations \cite{yexp}. The $Y(4260)$ was 
also observed in the 
$B^-\to Y(4260)K^-\to J/\Psi\pi^+\pi^-K^-$ decay \cite{babary2}, and CLEO
reported two additional decay channels: $J/\Psi\pi^0\pi^0$ and
$J/\Psi K^+K^-$ \cite{yexp}. 

The mass of the $Y(4260)$ is higher than the $D^{(*)}\bar{D}^{(*)}$
threshold, therefore, if it was a normal $c\bar{c}$ charmonium state, it 
should decay mainly into $D^{(*)}\bar{D}^{(*)}$. However, the
observed $Y$ state does not match the peaks in $e^+e^-\to D^{(*)\pm}D^{(*)
\mp}$ cross sections measured  by Belle \cite{belle5} and BaBar 
\cite{babar5,babar6}.
Besides, the $\Psi(3S),~\Psi(2D)$ and $\Psi(4S)$ $c\bar{c}$ states have 
been assigned to the well established $\Psi(4040),~\Psi(4160),~$ and 
$\Psi(4415)$ mesons respectively, and the prediction from quark models 
for the $\Psi(3D)$  state is 4.52 GeV. Therefore, the mass of the $Y(4260)$
is not consistent with any of the $1^{--}$ $c\bar{c}$ states 
\cite{review,Zhu:2007wz,kz}. 

There are many theoretical interpretations for the $Y(4260)$ such as
tetraquark state, hadronic molecule of 
$D_{1} D$, $D_{0} D^*$, $\chi_{c1} \omega$, $\chi_{c1} \rho$, $J/\psi f_0(980)$,
hybrid charmonium, charm baryonium, cusp, etc 
\cite{Brambilla:2014jmp,Hosaka:2016pey,review}. However, there are some
calculations, within the QCDSR approach that can not explain 
the mass of the $Y(4260)$ treating it as  a tetraquark state \cite{rapha},  
 as  a $D_{1} D$, $D_{0} D^*$ hadronic molecule \cite{rapha}, or  a  
$J/\psi f_0(980)$ molecular state \cite{Albuquerque:2011ix}. Therefore,  
as in the case of the $X(3872)$, in Ref.~\cite{Dias:2012ek}  the $Y(4260)$ 
was treated as a mixture of a $c\bar{c}$ state with a four-quark state: 
\beq
j_\mu(x)=\sin(\theta) \:j_\mu^{(4)}(x)+\cos(\theta) \:j_\mu^{(2)}(x),
\label{jmix}
\enq
where
\beq
j_\mu^{(4)}(x) = \frac{\epsilon_{abc} \epsilon_{dec}}{\sqrt{2}}
\Big[(q_a^T(x)C\gamma_5 c_b(x))(\bar{q}_d(x)\gamma_\mu\gamma_5 C\bar{c}_e^T(x))+
(q_a^T(x)C\gamma_5\gamma_\mu c_b(x))(\bar{q}_d(x)\gamma_5 C\bar{c}_e^T(x)) 
\Big],
\label{j4y}
\enq
and
\beq
j_\mu^{(2)}=\frac{1}{\sqrt{2}}\qq ~\bar{c}_a(x)\gamma_\mu c_a(x).
\enq

Varying the value of the mixing angle in the range $\theta=(53.0\pm0.5)^0$ it
was found  \cite{Dias:2012ek} that 
$m_{Y} = (4.26 \pm 0.13) ~ \mbox{GeV}$,
which is in a very good agreement with the experimental mass of the $Y(4260)$.

The width of the decay channel $Y(4260)\to J/\psi\pi\pi$, was also evaluated
in Ref.~\cite{Dias:2012ek} considering the same mixing angle and assuming 
that the two pions in the final
 state come from the $\sigma$ and $f_0(980)$ scalar mesons. The obtained 
value for the width is $\Gamma_{Y\to J/\psi\pi\pi} \approx (4.1\pm 0.6)$ MeV, 
which is much smaller 
than the total  experimental width: $\Gamma_{exp} \approx (95\pm 14)$ MeV
\cite{pdg}.

To compare the decay width into the $J/\psi\pi\pi$ channel with the
total width we have to consider other possible decay channels.
With the mixed current, the main decay channel of the   $Y(4260)$ should be 
into $D$ mesons, mostly due to the charmonium part of the current, but also 
from the tetraquark part through quark rearrangement.  Therefore, the total 
width of the $Y(4260)$ should be given by the sum of the partial widths of 
all these channels. Unfortunately, the QCDSR approach  does not allow the
evaluation of the decay channels involving $D$ mesons, 
since one can only 
use the QCDSR approach to study properties of the low-lying state. Therefore, 
the charmonium part of the current can only be used to study the 
decay of $J/\psi$. 

If  one considers the experimental upper limits, from the  BaBar 
\cite{babar6} and CLEO \cite{CroninHennessy:2008yi} collaborations, for 
the branching ratios
\beq
{\cal B}(Y(4260)\to X)\over{\cal B}(Y(4260)\to J/\psi\pi\pi),
\enq
where $X=D\bar{D},~D\bar{D}^*$ and $D^*\bar{D}^*$, one can see
that the width obtained in \cite{Dias:2012ek}, for the $J/\psi\pi\pi$ channel, is 
consistent 
with the total  experimental width of the $Y(4260)$. Therefore, they concluded 
that it is possibile to explain the $Y(4260)$ exotic state as
a mixed charmonium-tetraquark state.    

\subsection{$Z_c^+(3900)$}

From March to October of 2013 the BESIII collaboration reported the observation of
four charmonium charged states. 
The first one was the $Z_c^+(3900)$ \cite{Ablikim:2013mio}, observed almost 
at the same time by  the  Belle collaboration \cite{Liu:2013dau}, in 
the $M(\pi^\pm J/\psi)$ mass spectrum of the $Y(4260)\to J/\psi\pi^+\pi^-$ 
decay channel.  
The existence of this structure was promptly confirmed by 
the  authors of Ref.~\cite{Xiao:2013iha} using  CLEO-c data.   

Assuming $SU(2)$  symmetry, the mass obtained in QCDSR for the $Z_c$  coincides
with the one obtained  for the $X(3872)$. However, the
$Z_c(3900)$ decay width represents a challenge to theorists. While its mass 
is very close to the $X(3872)$ mass, which may be
considered its isosinglet partner, it has a much larger decay width. Indeed, 
while the $Z_c(3900)$ decay width is in the range $40-50$ MeV, the $X(3872)$ 
width is  smaller than  $ 1.2 $ MeV.  
This difference can be attributed to  the fact that the $X(3872)$
may contain a significant $|c \bar{c} \rangle$ component \cite{x24}, which is
absent in the $Z_c(3900)$.  As pointed out in Ref.~\cite{zhao}, this 
would also explain why the $Z_c$ has not been observed in $B$ decays. 

According to the experimental observations, the $Z_c(3900)$ decays into 
$J/\psi \,  \pi^+$ with a relatively large decay width. This is unexpected 
for a $D^* - \bar{D}$  molecular state, in which the distance between the  
$D^*$ and the $ \bar{D}$ is large. This decay must involve the exchange of a 
charmed meson, which is a short range process and hence unlikely to occur in 
large systems. In Ref.~\cite{namit}  it was shown that, in order to reproduce 
the measured width, the effective radius must be $\langle r_{eff} \rangle 
\simeq 0.4$ fm. This size scale is small and pushes the molecular picture to 
its limit of validity. In another work \cite{hammer}, the new state
was treated as a charged $D^* - \bar{D}$ molecule and the authors explored 
its  electromagnetic structure, arriving 
at the conclusion that its charge radius is of the order of  $\langle r^2 
\rangle \simeq 0.11$ fm$^2$. 
Taking this radius as a measure of the spatial size of the state, 
we conclude that it is more compact than a $J/\psi$, for which
$\langle r^2 \rangle \simeq 0.16$ fm$^2$.  In Ref.~\cite{Dias:2013xfa} the 
combined results of refs.~\cite{namit} and \cite{hammer} were taken  as 
an indication that the $Z_c$ is a compact object, which may be better 
understood as a quark cluster, such as a tetraquark.  Moreover, the 
$Z_c(3900)$ was interpreted  as the isospin 1 partner of the $X(3872)$, as 
the charged state predicted in Ref.~\cite{maiani}. Therefore, the quantum 
numbers for the neutral state in the isospin multiplet were assumed to be 
$I^G(J^{PC})=1^+(1^{+-})$. The interpolating field for $Z_c^+(3900)$ used in 
Ref.~\cite{Dias:2013xfa} is given by Eq.~(\ref{cur-di})  with the plus sign 
changed to a minus sign. The three-point QCDSR was used to evaluate 
the coupling constants in the vertices $Z_c^+(3900)J/\psi\pi^+$,  
$Z_c^+(3900)\eta_c\rho^+$,   $Z_c^+(3900) D^+ \bar{D^*}^0 $ 
and   $Z_c^+(3900) \bar{D^0} {D^*}^+ $.  

\begin{figure}[h]
%\leavevmode
\sidecaption
\includegraphics[width=5cm,clip]{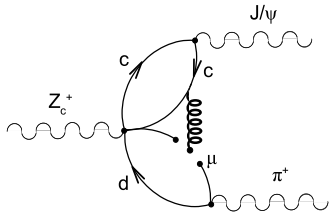}
%\centerline{\epsfig{figure=diagramcc.eps,height=40mm}}
%,width=70mm,angle=0}}
\caption{CC diagram which contributes to the OPE side of the sum rule.}
\label{fig1}
\end{figure} 

In the case of the $Z_c^ +\to J/\psi\pi^ +$ decay,  the generic 
decay diagram in terms of quarks has two ``petals'', one associated with the 
$J/\psi$ and the other with the $\pi^ +$. Among the possible diagrams, there 
are two distinct subsets.  Diagrams with no gluon exchange between the petals 
and, therefore, no color exchange between the two final mesons in the decay. 
If there is no color exchange, the final state containing two color singlets 
was already present in the initial state. This happens  because, although the 
initial current,  Eq.~(\ref{cur-di}), has a non-trivial color structure, it 
can be rewritten as a sum of molecular type currents with trivial color 
configuration through a Fierz transformation. To avoid this problem we 
consider in the OPE side only the diagrams with non-trivial color structure, 
as the one shown in Fig.~1.  
This type of diagram represents the case where the $Z_c^ +$ is a 
genuine four-quark state with a complicated color structure. These diagrams 
are called color-conected (CC).

Here  only color-connected 
diagrams were considered, since the $Z_c(3900)$ is expected to be a genuine 
tetraquark state with a non-trivial color structure. The obtained couplings, 
with the respective decay widths, are given in Table 2. 
A total width of $\Gamma = (63.0 \pm 18.1)$ MeV was found for the  
$Z_c(3900)$,  
in good agreement with the two experimental values:
$\Gamma=(46\pm 22)$ MeV from BESIII \cite{Ablikim:2013mio}, and
$\Gamma=(63\pm35)$ MeV from BELLE \cite{Liu:2013dau}.

\begin{table}[h]
\begin{center}
\caption{ Coupling constants and decay widths in  different channels}
\label{tab2}       
\begin{tabular}{|c|c|c|} \hline
Vertex & coupling constant (GeV) & decay width (MeV)\\
\hline
$Z_c^+(3900)J/\psi\pi^+$ & $3.89\pm0.56$ & $29.1\pm8.2$ \\
\hline
 $Z_c^+(3900)\eta_c\rho^+$ & $4.85 \pm 0.81$ & $27.5\pm8.5$ \\
\hline
 $Z_c^+(3900) D^+ \bar{D^*}^0 $ & $2.5 \pm 0.3$ & $3.2 \pm 0.7$ \\
\hline
 $Z_c^+(3900) \bar{D^0} {D^*}^+ $ & $2.5 \pm 0.3$ & $3.2 \pm 0.7$ \\ 
\hline
\end{tabular}
\end {center}
\end{table}

From the results in Table 2 it is possible to evaluate the ratio
\begin{equation}
{\Gamma(Z_c(3900) \to D\bar{D}^*)\over
\Gamma(Z_c(3900) \to\pi J/\psi)}=0.22 \pm 0.12. 
\label{ratio}
\end{equation}

\subsection{$Z^+_c (4025)$}

Soon after the $Z_c^+(3900)$ observation, the BESIII collaboration reported 
the observation 
of other three charged states: $Z_c^+(4025)$ \cite{Ablikim:2013emm}, 
$Z_c^+(4020)$ \cite{Ablikim:2013wzq}    and $Z_c^+(3885)$ 
\cite{Ablikim:2013xfr}.
Up to now it is not clear if the states $Z_c^+(3900)$-$Z_c^+(3885)$
and the states $Z_c^+(4025)$-$Z_c^+(4020)$ are the same states seen in 
different decay channels, or if they are independent states. 

In the case of $Z_c^+(4025)$, a study of the reaction $ e^+ e^- \to(D^* 
\bar{D^*})^{\pm} \pi^{\mp} $ was performed by the BESIII Collaboration at 
$\sqrt{s} = 4.26$ GeV and a peak was  seen in the $ (D^*  \bar{D^*})^{\pm}  $ 
invariant mass distribution just about $10$ MeV above the threshold. 
\cite{Ablikim:2013emm}. 
The authors assume in the paper that the  $(D^*\bar{D^*})^{\pm}$
 pair is created in a S-wave and then the $Z_c^+(4025)$  
must have $J^P = 1^+$  to match, together with the pion, the quantum numbers 
$J^P = 1^-$ of the virtual photon from the $e^+ e^-$ pair. However, they also 
state that the experiment does not exclude other spin-parity assignments. 

Many theoretical papers were devoted to understand these new states. 
In Ref.~\cite{hidalgo}, assuming the $X(3872)$  to be a $D \bar{D^*}$ 
molecule, the authors found a series of new hadronic molecules, including the 
$Z_c^+(3900)$ and the $Z_c^+(4025)$. They  would correspond to bound states 
(with uncertainties of about $50$ MeV in the binding) of $D\bar{D^*}$ and 
$D^*\bar{D^*}$  respectively, with quantum numbers $I(J^{P}) = 1(1^{+})$.  
Remarkably, even with uncertainties, these states always appear  in the bound 
region. In refs.~\cite{chen,cui}, using QCDSR and assuming a
structure of $D^* \bar{D^*}$, the authors obtained a possible $I(J^P ) = 1(1^+)$
 state compatible with the $Z_c^+(4025)$, but with  $\sim 250$ MeV 
uncertainty in the energy. In Ref.~\cite{qtang}, using a tetraquark current
and QCDSR, a state with $I(J^P ) = 1(2^+)$ compatible with $Z_c(4025)$ was 
obtained, once again with a large error in the energy of $190$ MeV.
In Ref.~\cite{sunzhu}  the new $Z_c$ states were 
investigated from a different perspective and, using pion exchange, a 
$D^* \bar{D^*}$ state with 
$I(J^P ) = 1(1^+)$ compatible with the $Z_c(4025)$ was obtained.

 A moleculelike  picture for  $Z_c(4025)$  seems to be quite plausible since 
its mass is merely 8 MeV away from  the $\bar{D}^{*0} D^{*+}$ threshold. 
In Ref.~ \cite{khetonn}, a study of the 
$D^* \bar{D^*}$ system has also been done within QCD sum rules, using a
interpolating current corresponding to the  $\bar{D}^{*0} D^{*+}$ molecule. The
 idea was to test if the $Z_c(4025)$ could be interpreted as a  $1^+$ or $2^+$ 
resonance of the  $\bar{D}^{*0} D^{*+}$ system. The $0^+$ assignment is ruled 
out for $Z_c(4025)$ by spin-parity conservation for the
  $e^+ e^- \to \left( D^* \bar{D}^* \right)^\pm \pi^\pm$ process. The tensor
interpolating current used in \cite{khetonn} was:
\begin{equation}\label{j}
j_{\mu \nu} (x) = \left[ \bar{c}_a(x) \gamma_\mu u_a(x)\right]\left[\bar{d}_b(x)
 \gamma_\nu c_b(x)\right],
\end{equation}
where $a,b$ denote the color indices. The corresponding two-point  correlation 
function is:
\begin{equation}
\Pi_{\mu \nu \alpha \beta} (q^2) = i \int  d^4x e^{iqx} \langle 0 \mid T \left[
 j_{\mu \nu} (x) j^\dagger_{\alpha \beta} (0) \right] \mid 0 \rangle .\label{Pi}
\end{equation}
The $0^+$, $1^+$ and $2^+$ components of the correlation function written in 
Eq.~(\ref{Pi}) can be 
obtained by using the following projectors
\begin{align} 
\mathcal{P}^{(0)}&=\frac{1}{3}\Delta^{\mu\nu}\Delta^{\alpha\beta},\nonumber\\
\mathcal{P}^{(1)}&=\frac{1}{2}\left(\Delta^{\mu\alpha}\Delta^{\nu\beta}-
\Delta^{\mu\beta}\Delta^{\nu\alpha}\right),\label{proj}\\
\mathcal{P}^{(2)}&=\frac{1}{2}\left(\Delta^{\mu\alpha}\Delta^{\nu\beta}+
\Delta^{\mu\beta}\Delta^{\nu\alpha}\right)-\frac{1}{3}\Delta^{\mu\nu}
\Delta^{\alpha\beta},\nonumber
\end{align}
where $\Delta_{\mu\nu}$ is defined in terms of the metric tensor, 
$g^{\mu\nu}$, and the four momentum $q$ of the correlation function as
 \begin{align}
\Delta_{\mu\nu}\equiv -g_{\mu\nu}+\frac{q_\mu q_\nu}{q^2}.\label{Delta}
 \end{align}
In the three cases a state with mass $3950 \pm 100$ MeV was found \cite{khetonn}.
 The central value of the mass 
of these states is more in line with the results of Refs.~\cite{raquel,oset}, 
although with the error bar, they could as well be related to a resonance.

Bumps close to the threshold of a pair of particles should be treated with 
caution. Sometimes they are identified as new particles, but they can also be 
a reflection of a resonance below threshold. Further 
examples of this phenomenon  may be found in Ref.~\cite{tokhenno14}, where 
the theory of  $ D^* \bar{D^*}  $   interactions is reviewed and it 
is pointed out that a  $ (D^* \bar{D^*})  $  state with a mass above the 
threshold is very difficult to support. In particular, in Ref.~\cite{oset} 
it was found that there is only one bound state of $ (D^* \bar{D^*})  $ 
in $I^G = 1^-$, with quantum numbers $J^{PC} = 2^{++}$  with a mass around 
$3990$ MeV and a width of about $100$ MeV. Both mass and width are compatible
with the reanalysis of data carried out in \cite{tokhenno14}. Therefore,
we can conclude that such $J^{P} = 2^{+}$ $D^{*}\bar{D}^*$ bound state provides 
a natural explanation for the state observed in  \cite{Ablikim:2013emm}.

An argument against the existence of a new resonance above the threshold is the 
fact that if the state were a $J^P = 1^+$  produced in S-wave, as assumed in the 
experimental work, it would easily decay into $J/\psi \pi$  exchanging a $D$ 
meson in the t-channel. This is also the decay channel of
the $Z_c(3900)$, which would then have the same quantum numbers as
the state claimed in Ref.~\cite{Ablikim:2013emm}. However, while a peak is 
clearly seen in the $J/\psi \pi$ invariant mass distribution in the case of the 
$Z_c(3900)$, no trace of a peak is seen around $4025$ MeV in spite of using the 
same reaction and the same $e^+ e^- $ energy.

\subsection{$X^+ (5568)$} 

This year the  D0  Collaboration   has  announced the observation  of  
a  new state in the $B_{s}^0\pi^\pm$ mass 
spectrum, the $X^\pm(5568)$ \cite{D0:2016mwd}.The $X(5568)$ would be a very 
important addition to the list of undoubtedly exotic mesons, since its wave 
function consists of four different flavors: $u$, $b$, $d$ and $s$ quarks. 
However,  the LHCb Collaboration has not confirmed the observation of the 
$X(5568)$ \cite{Aaij:2016iev}, since in their analysis no structure 
is found in the $B_{s}^0\pi^\pm$ mass spectrum from the  $B_{s}^0\pi^+$ 
threshold up to $M_{B_s^0\pi^+}\leq 5.7$ GeV.  

The announcement of the exotic state $X(5568)$  stimulated the theoretical 
interest and  several  theoretical works have been done to investigate the 
properties of such state. There are studies based on QCDSR, 
quark models, coupled channel analysis  and more general arguments. In some of 
these studies it was not possible to explain the reported properties of  the 
$X(5568)$ neither as a molecule nor as a tetraquark state
\cite{Burns:2016gvy,Guo:2016nhb,Wang:2016tsi,Ali:2016gdg,Zanetti:2016wjn,Chen:2016npt}. 
However, in many other calculations it was possible to explain
the reported properties 
~\cite{Agaev:2016mjb,Wang:2016mee,Tang:2016pcf,He:2016yhd,Liu:2016ogz,Stancu:2016sfd,Albaladejo:2016eps,Chen:2016mqt}. 

In Ref.~\cite{Zanetti:2016wjn} we used a $(I)J^P=(1) \, 0^+$ 
scalar-diquark scalar-antidiquark tetraquark current for the $X^+(5568)$:
\beq
j_S={\epsilon_{abc}\epsilon_{dec}}(u_a^TC
\gamma_5s_b)(\bar{d}_d\gamma_5C\bar{b}_e^T),
\label{int}
\enq
where $a,~b,~c,~...$ are colour indices and $C$ is the charge conjugation
matrix. 
Our  study indicates that although  it is possible to obtain a stable mass in 
agreement with the state found by the D0 collaboration, a more restrictive
analysis (simultaneous requirement of the OPE convergence and the dominance of 
the pole on the phenomenological side) leads to a higher mass. In particular,
considering condensates up to dimension 8 we get~\cite{Zanetti:2016wjn}:
\beq 
m_X=(6.39\pm0.10)\GeV,   
\label{result2}
\enq 
 which  is  not in agreement  with  the experimental   mass   of   the    
$X(5568)$   determined   by   the   D0  Collaboration \cite{D0:2016mwd},
leading us to conclude  that  the  $X(5568)$  state
can not be represented by the scalar tetraquark current.

Clearly, more  analysis are required to clarify this situation from the 
experimental side as well as from the theoretical side.

\section{Conclusions}

Here we have reported the masses of some of these $X$, $Y$ and $Z$ 
states, using the QCDSR approach. In some cases a tetraquark configuration was 
favored, as for the $Z_c^+(3900)$, and in some other cases a molecular 
configuration was favored, like the  $Z_c^+(4025)$. 
In the case of the $X(3872)$ (and also $Y(4260)$) we found that it is only possible to 
explain all the available experimental data if it is a mixed state with
charmonium and four-quark components, and in the case of the $X^+(5568)$ it
was not possible to describe it as a tetraquark state.

The most important message from the experimental program carried out by the 
 BaBar, Belle, CLEO-c, CDF, \DZero, BESIII, LHCb and CMS 
Collaborations is that definitely there is something really 
new happening in the charmonium spectroscopy. The program  started in 2003 with 
the measurement of the $X(3872)$. There is no doubt in the community that the 
$X(3872)$ structure is more complex than a simple $c\bar{c}$ state. However, we 
can say that the confirmation of the observation of the $Z^+(4430)$ by the LHCb 
Collaboration together with the measurements of the $Z^+_ c(3900)$, which was 
measured by BESIII and confirmed by other groups, reinforced our belief that we 
are observing multiquark states.  

 In the next years it is important: i) from the experimental side 
to determine the quantum numbers of all these states and eliminate the 
suspicion that some of them could be mere threshold effects and not real 
particles.  ii) from the theoretical side to focus on the 
calculation of the decay widths in all the different approaches, since, as we 
have discussed, the masses are easily obtained by different methods and they 
are not sufficient to discriminate between different theoretical models.  

 \subsection*{Acknowledgments}
 \noindent
%%%%%%%%%%%%%%%%
%\acknowledgments
This work has been supported by  CNPq and FAPESP-Brazil.

\end{document}